\documentstyle[prl,aps,epsfig,calc,graphicx]{revtex}         % 2 col mode

\begin{document}
\draft               % preprint mode
\newcommand{\li}{$^6$Li }
\newcommand{\na}{$^{23}$Na }
% 2 col mode:
\twocolumn [\hsize\textwidth\columnwidth\hsize\csname
@twocolumnfalse\endcsname %\vspace{-5mm}

% Front matter --------------------------------------------------------------

\title{Decay of an ultracold fermionic lithium gas near a Feshbach resonance}
\author{K. Dieckmann, C. A. Stan, S. Gupta, Z. Hadzibabic, C. H. Schunck, and W. Ketterle}
\address{Department of Physics, MIT-Harvard Center for Ultracold
Atoms, and Research Laboratory
of Electronics, \\
MIT, Cambridge, MA 02139}
\date{\today}
\maketitle

\begin{abstract}
We studied the magnetic field dependence of the inelastic decay of
an ultracold, optically trapped fermionic $^6$Li gas of different
spin compositions. The spin mixture of the two lowest hyperfine
states showed two decay resonances at $550\,$G and $680\,$G,
consistent with the predicted Feshbach resonances for elastic
s-wave collisions. The observed lifetimes of several hundred
milliseconds are much longer than the expected time for Cooper
pair formation and the phase transition to superfluidity in the
vicinity of the Feshbach resonance.
\end{abstract}

\vskip1pc ]

% Main matter ---------------------------------------------------------------

\narrowtext

% Introduction -------------------------------------------------------------
Interactions between atoms can be strongly modified by tuning
magnetic fields to Feshbach resonances where a molecular state has
the same energy as the colliding atoms.  This mechanism has been
used to dramatically alter the properties of ultracold bosonic
gases \cite{Inou98,Cour98,Vule99,Corn00}. For degenerate Fermi
gases, such control over the interaction strength is crucial in
the search for a superfluid phase transition. For dilute Fermi
gases, the predicted phase transition occurs at temperatures that
are experimentally not accessible \cite{Legg80}, unless the
scattering length is resonantly enhanced. In this case, as was
pointed out by \cite{Houb97,Timm01,Holl01,Ohas02,Mils02}, the
transition temperature can be comparable to the temperatures
achieved in current experiments
\cite{dema01,trus01,schr01,Gran02,Hadz02,roat02}.

Promising candidates for an experimental observation of fermionic
superfluidity are $^6$Li and $^{40}$K. For an optically trapped
mixture of two spin states of $^{40}$K, a Feshbach resonance has
been observed by measuring the thermalization time of the gas
\cite{Loft01}. For an optically trapped spin mixture of the two
lowest Zeeman states of $^6$Li, a wide s-wave Feshbach resonance
has been predicted first by \cite{Houb98}. Experiments with $^6$Li
have so far only observed a magnetic field dependence of the
elastic cross section far away from the predicted resonance
\cite{Gran02}. Near Feshbach resonances, the enhancement of the
scattering length is usually accompanied by enhanced inelastic
collisions which lead to rapid trap loss. This signature was used
to identify Feshbach resonances in bosonic gases
\cite{Sten99,Robe00,Chin00}. However, inelastic losses have also
posed a severe limitation for experiments near Feshbach
resonances, in particular at high atomic densities. The superfluid
phase transition for fermions will only be observable if the time
for the formation of Cooper pairs is shorter than the decay time
of the gas. For fermions, inelastic decay in the s-wave channel
can be suppressed due to the Pauli exclusion principle. However,
even in the zero-temperature limit the kinetic energy of the cloud
is of the order of the Fermi energy. Therefore, inelastic
collisions for higher partial waves are expected to limit the
lifetime of the gas.

This letter is the first report on the study of inelastic
collisions in a fermionic system near Feshbach resonances. We have
observed resonant magnetic field dependent inelastic decay of an
ultracold, optically trapped spin mixture of $^6$Li.

% Preparation of the spin mixture ------------------------------------------
The ultracold lithium samples were prepared by sympathetic cooling
of \li by \na as described previously \cite{Hadz02}. The remaining
$^{23}$Na atoms were removed from the magnetic trap by rf induced
spin-flips to untrapped states. This typically produced $3\times
10^5$ lithium atoms in the $|1/2,-1/2\rangle$ state at a
temperature of $400\,$nK, equal to half the Fermi temperature. The
atoms were transferred into an optical trap formed by a single far
detuned beam with up to $1\,$W of power at $1064\,$nm. The beam
had a $14\,\mu$m waist and was aligned horizontally along the
symmetry axis of the magnetic trap. This generated a $175\,\mu$K
deep trapping potential, with $12\,$kHz radial and $200\,$Hz axial
trapping frequencies. Prior to the transfer, the cloud was
adiabatically decompressed in the radial direction during $1\,$s
to improve the spatial overlap with the optical trap. After this
stage, the trap frequencies in the magnetic trap were $149\,$Hz
radially and $26\,$Hz axially. We then adiabatically ramped up the
power of the optical trap over $500\,$ms. Subsequently, the
magnetic trapping fields were ramped down in $100\,$ms, leaving a
$1.5\,$G guiding field along the trap axis. After the transfer,
the cloud contained $3\times 10^5$ atoms at $3\times
10^{13}\,$cm$^{-3}$ peak density and $22\,\mu$K temperature, close
to the $21\,\mu$K Fermi temperature. We attribute the rise in
temperature relative to the Fermi temperature to residual
excitations during the transfer into the optical trap. (We often
observed axial oscillations of the cloud after the transfer.)

We studied inelastic decay for three different spin compositions
of the cloud. The lithium atoms were either trapped purely in the
lowest ($|1\rangle$), or the second to lowest ($|2\rangle$) energy
state, or in a $50\%-50\%$ mixture of these two Zeeman states. At
low magnetic fields, the states $|1\rangle$ and $|2\rangle$
correspond to the $|F,m_F\rangle=|1/2,+1/2\rangle$ and
$|1/2,-1/2\rangle$ states, respectively. A full transfer
$|1/2,-1/2\rangle\rightarrow |1/2,+1/2\rangle$ was done at low
magnetic field by applying a $1\,$s rf-driven adiabatic passage
between the two states, which was $>95\%$ complete. The spin
mixture was produced by a faster, non-adiabatic rf-sweep of
$200\,$ms duration. The population of the states was analyzed by
applying a $7\,$G/cm magnetic field gradient along the trap axis
with a $6.5\,$G offset field in the center, and reducing the
strength of the optical confinement. This resulted in full spatial
separation of the two spin states in the optical trap. Resonant
absorption imaging was used to determine the atom number in each
of the spin states. Using a full transfer we compared the
absorption cross sections for circularly polarized light for the
two spin states and found a ratio of $1:1.2$. Taking this into
consideration, we were able to control the relative population of
the spin states by rf-sweeps with an accuracy of $\pm 4\,\%$.

% Feshbach scan experiment -------------------------------------------------
In order to study the decay of the cloud near the Feshbach
resonance, predicted to occur at about $800\,$G \cite{Houb98}, we
applied magnetic fields up to $900\,$G using the anti-bias coils
of the cloverleaf magnetic trap \cite{Mewe96}. The magnetic field
strength was calibrated in two independent ways to $2\%$ accuracy.
For calibration of magnetic fields up to $100\,$G, we loaded \na
into the optical trap and drove rf transitions to magnetically
untrapped states. Resonances were observed by measuring the
remaining atom number after recapture into the magnetic trap. As a
second method at about $700\,$G, we used direct absorption imaging
of \li in the optical trap in the presence of higher magnetic
fields. The magnetic field values were then derived from the
frequency shifts of the observed resonances from the lithium D2
line. We also verified that drifts of the magnetic field during
the pulses, occurring from thermal expansion of the coils due to
the high current load, were negligible.

We measured the magnetic field dependence of the decay by
measuring the atom number at two different times, $50\,$ms and
$500\,$ms after switching on the magnetic field within about
$4\,$ms. For measuring the remaining atom number, the magnetic
field was rapidly switched off within $100\,\mu$s, and the cloud
was probed by absorption imaging at low magnetic field.
Normalizing the number at long time to the number at short time
made the measurement less sensitive to atom number drifts and
initial losses from the optical trap. These losses can occur due
to the sloshing motion of the cloud and due to initial
evaporation.

\begin{figure}[tb]\hspace{-4mm}
\epsfxsize=80mm{\epsfbox{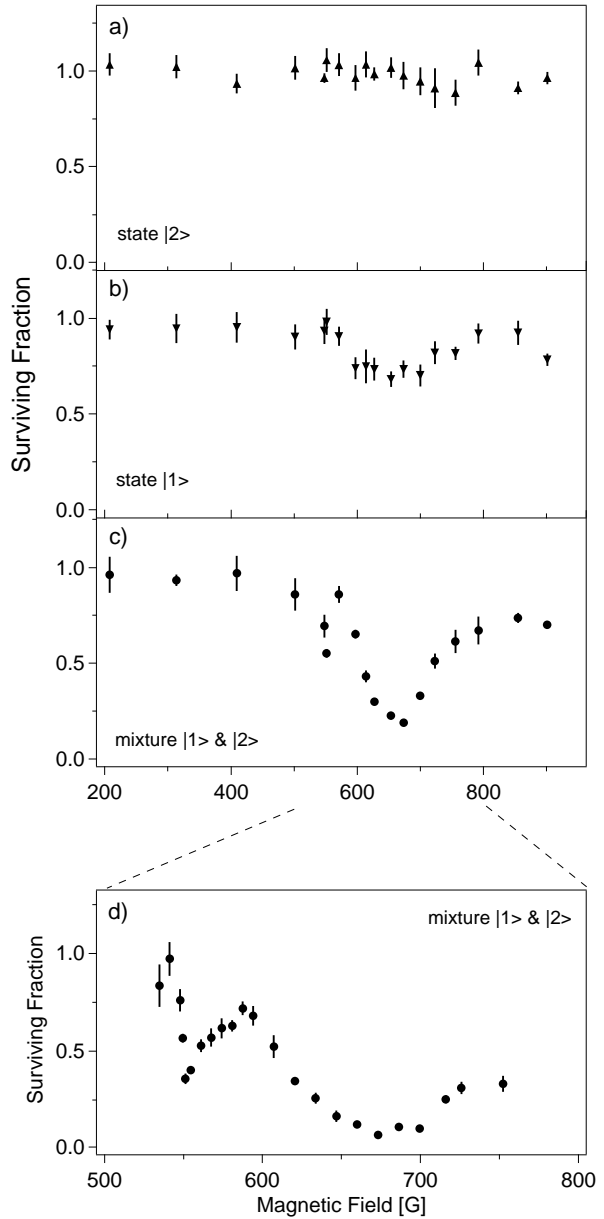}}
\caption{\label{fig:resonancescan} Magnetic field dependence of
inelastic decay in clouds of fermionic $^6$Li. The fraction of
atoms remaining after the $500\,$ms magnetic field pulse is shown
for different spin compositions of the cloud. a) For the state
$|2\rangle$, no significant loss was observed. b) The
energetically lowest state $|1\rangle$ exhibits a weak decay
resonance at $\approx680\,$G. c) The $50\%-50\%$ mixture of two
spin states shows two decay resonances, at $550\,$G and $680\,$G.
d) The two resonances are shown with higher density of data points
and for $2\,$s magnetic field pulses. Each data point represents
an average of three measurements.}
\end{figure}

For the cloud purely in state $|2\rangle$, we observed no
significant decay over the entire range of magnetic fields, as can
be seen in Figure~\ref{fig:resonancescan}, a). This also confirmed
that during the measurement interval, one-body decay (e.g. due to
collisions with particles from the background gas) was negligible.

The surviving fraction of the mixture is shown in
Figure~\ref{fig:resonancescan}, c). No significant decay was
observed at low magnetic fields. At higher magnetic field, we
found two decay resonances. A strong resonance occurred at
$680\,$G with considerable losses over a range of approximately
$100\,$G. At even higher magnetic fields, the decay persisted at a
weaker but constant level. In a more detailed scan, shown in
Figure~\ref{fig:resonancescan}, d), a second, much weaker and
narrower resonance was found at $550\,$G, with an
approximate width of $20\,$G. The weaker resonance became more
pronounced after $2\,$s of dwell time in the magnetic field,
whereas the stronger resonance showed ``saturation'' broadening.

\begin{figure}[htb]\hspace{-4mm}
\epsfxsize=80mm{\epsfbox{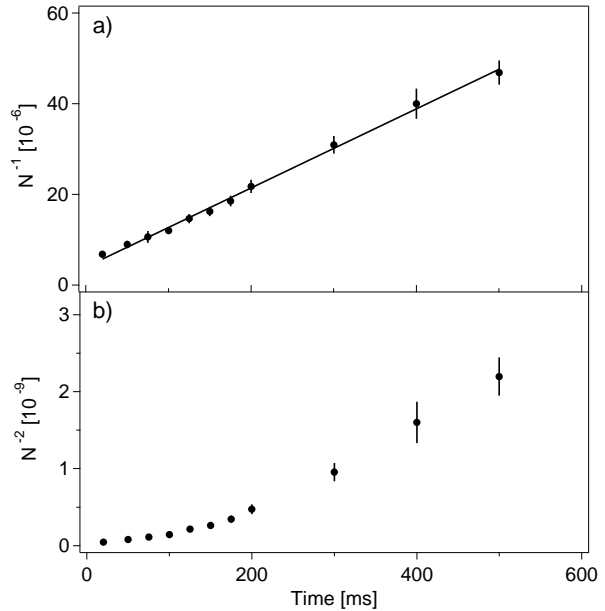}}
\caption{\label{fig:decaycurves} Decay of the atom number at
$680\,$G. a) The data plotted as $1/N$ show a linear time
dependency, consistent with two-body decay. b) The same data
plotted as $1/N^2$ clearly show non-linear dependency. For the
resonance at $550\,$G, the comparison of least square fits also
revealed consistency with a two-body decay.}\vspace{2mm}
\end{figure}

% decay experiment and fits -------------------------------------------------
We also measured the time evolution of the atom number at the two
resonances. For a two-body (three-body) process the loss rate of
atoms $\dot{N}$ is proportional to $N^2$ ($N^3$), where N is the
number of trapped atoms. The decay curves at $680\,$G are shown in
Figure~\ref{fig:decaycurves}. At both resonances we found that the
values for $1/N$ showed a linear dependence on time,
characteristic for a two-body process. In order to test for
three-body decay we plotted the same data as $1/N^2$. The
nonlinear behaviour is not compatible with a simple three-body
decay process.

Another experimental observation is the almost complete
disappearance of the mixed cloud in
Figure~\ref{fig:resonancescan}, d). A resonantly enhanced
three-body process would involve two atoms of opposite spin
colliding, and a third in either of the spin states. Starting with
a $50\%-50\%$ mixture, the decay would stop when all atoms in
state $|1\rangle$ (or in state $|2\rangle$) are used up.
Therefore, three-body decay can only be consistent with the
observation of complete disappearance, if the decay rate does not
depend on the spin state of the third particle. In case of
strongly different rates for the two spin orientations, the
surviving fraction could not drop below $25\%$.

% Interpretation -----------------------------------------------------------
With the observation of two resonances and the position of the
strongest decay of the main resonance deviating from the
theoretical prediction \cite{Houb98}, the question arises whether
the observed decay of the spin mixture can be interpreted as a
signature of the Feshbach resonance for elastic s-wave collisions.
After the submission of this paper new improved theoretical
calculations exhibited a second narrow Feshbach resonance for
elastic collisions in the s-wave channel at $550\,$G
\cite{Hara02}, in good agreement with the position of the narrow
decay resonance. The predicted magnetic field for the main
resonance is $860\,$G. However, due to the huge width of the
resonance it seems possible that the decay observed at $680\,$G is
related to this s-wave resonance.

The measured decay curves suggest a two-body type of decay. Due to
the Pauli exclusion principle dipolar relaxation is not possible
in the s-wave channel \cite{Stoo88}. Dipolar relaxation in the
p-wave channel is possible, as even in the zero-temperature limit
the kinetic energy of the cloud is of the order of the Fermi
energy, and collisions in the p-wave channel do not completely
freeze out. However, no occurrences of resonances in the dipolar
decay are theoretically predicted \cite{Kokk02}.

Therefore, it is most likely that the observed decay is a
signature of the Feshbach resonances for the elastic collisions,
resulting in enhanced three-body decay. At present no exact
theoretical description for the three-body decay mechanism of
fermions near a Feshbach resonance is available. Three-body decay
is not supported by the measured decay curves. However, one
possibility is that
the decay curve is affected by a change in temperature. An
accurate measurement of the temperature was difficult due to
technical reasons and a low signal to noise ratio, as the
absorption signal drops significantly during the decay. If the
sample had cooled down during the decay (e.g. due to an energy
dependence of the loss rate) it could speed up the decay in a way
that three-body loss results in a decay curve similar to a curve
for two-body losses at constant temperature. Another possibility
for the deviation from a three-body decay curve would be heating
due to three-body recombination followed by trap loss due to
evaporation, or other processes involving secondary collisions
\cite{Petr02}. It should be noted that the observed resonances do
not resemble the predicted magnetic field dependence for elastic
collisions \cite{Houb98}. Therefore, our decay data cannot be
explained by elastic collisions leading to evaporation.

We also observed resonant decay at $680\,$G of a cloud purely in
state $|1\rangle$, as shown in Figure~\ref{fig:resonancescan}, b).
The fact that this resonance is at the same magnetic field as for
the mixture suggests that the observed loss is due to a
contamination of the cloud with atoms in state $|2\rangle$. For
three-body of decay, our measured $>95\%$ purity of the
preparation of state $|1\rangle$ allows for a maximum of $15\%$
decay of the cloud, compared to the measured $21\,$\%. Further
measurements are needed to investigate whether there is an
enhancement of losses by secondary collisions, or whether there is
a decay mechanism for atoms purely in state $|1\rangle$.

% Conclusions --------------------------------------------------------------
In conclusion, we observed two decay resonances for the $^6$Li
spin mixture and one resonance in the lowest spin state. Comparing
our observations with recent theoretical calculations which
exhibit two s-wave Feshbach resonances suggests that the observed
decay is a signature of those resonances. Even on resonance, the
observed decay happened on a time scale longer than the trap
oscillation time, the time for elastic collisions, and the
expected sub-millisecond time needed for the formation of Cooper
pairs \cite{Houb99,Timm01b}. Therefore, the $^6$Li system is well
suited for the study of an interacting Fermi gas in the vicinity
of an elastic Feshbach resonance, in particular for the search for
the phase transition to a superfluid state.

% End matter ----------------------------------------------------------------

This research was supported by NSF, ONR, ARO, NASA, and the David
and Lucile Packard Foundation. C.~H.~S. acknowledges the support
of the Studienstiftung des deutschen Volkes.

{\it Note added:} After the submission of this paper several
groups reported related results. Measurements of the elastic cross
section near the zero crossing associated with the Feshbach
resonance have recently been performed by \cite{Hara02,Joch02}.
Inelastic decay of $^{6}$Li fermionic clouds near the Feshbach
resonance was recently also observed in the groups of J.~E. Thomas
\cite{Hara02}, and C. Salomon, and for $^{40}$K in the group of
D.~S. Jin.

%\bibliographystyle{prsty}
%\bibliography{resonantdecayBib}

\begin{thebibliography}{10}

\bibitem{Inou98}
S. Inouye, M.~R. Andrews, J. Stenger, H.-J. Miesner, D.~M.
Stamper-Kurn, and W.
  Ketterle, Nature (London) {\bf 392},  151  (1998).

\bibitem{Cour98}
P. Courteille, R.~S. Freeland, D.~J. Heinzen, F.~A. van Abeelen,
and B.~J.
  Verhaar, Phys. Rev. Lett. {\bf 81},  69  (1998).

\bibitem{Vule99}
A.~J. Kerman, V. Vuleti\'{c}, C. Chin, and S. Chu, Phys. Rev.
Lett. {\bf 82},
  1406  (1999).

\bibitem{Corn00}
S.~L. Cornish, N.~R. Claussen, J.~L. Roberts, E.~A. Cornell, and
C.~E. Wieman,
  Phys. Rev. Lett. {\bf 85},  1795  (2000).

\bibitem{Legg80}
A.~J. Leggett, J. Phys. (Paris) {\bf 41},  C7  (1980).

\bibitem{Houb97}
M. Houbiers, R. Ferwerda, H.~T.~C. Stoof, W.~I. McAlexander, C.~A.
Sackett, and
  R.~G. Hulet, Phys. Rev. A {\bf 56},  4864  (1997).

\bibitem{Timm01}
E. Timmermans, K. Furuya, P.~W. Milonni, and A.~K. Kerman, Phys.
Lett. A {\bf
  285},  228  (2001).

\bibitem{Holl01}
M.~J. Holland, S.~J.~J.~M.~F. Kokkelmans, M.~L. Chiofalo, and R.
Walser, Phys.
  Rev. Lett. {\bf 87},  120406  (2001).

\bibitem{Ohas02}
Y. Ohashi and A. Griffin,  Rev. Lett. {\bf 89},  130402  (2002).

\bibitem{Mils02}
J.~N. Milstein, S.~J.~J.~M.~F. Kokkelmans, and M.~J. Holland,
Phys. Rev. A {\bf 66},  043604  (2002).

\bibitem{dema01}
B. DeMarco, S.~B. Papp, and D.~S. Jin, Phys. Rev. Lett. {\bf 86},
5409
  (2001).

\bibitem{trus01}
A.~G. Truscott, K.~E. Strecker, W.~I. McAlexander, G.~B.
Partridge, and R.~G.
  Hulet, Science {\bf 291},  2570  (2001).

\bibitem{schr01}
F. Schreck, L. Khaykovich, K.~L. Corwin, G. Ferrari, T. Bourdel,
J. Cubizolles,
  and C. Salomon, Phys. Rev. Lett. {\bf 87},  080403  (2001).

\bibitem{Gran02}
S.~R. Granade, M.~E. Gehm, K.~M. O'Hara, and J.~E. Thomas, Phys.
Rev. Lett. {\bf 88},  120405  (2002).

\bibitem{Hadz02}
Z. Hadzibabic, C.~A. Stan, K. Dieckmann, S. Gupta, M.~W.
Zwierlein, A.
  G{\"o}rlitz, and W. Ketterle, Phys. Rev. Lett. {\bf 88},  160401  (2002).

\bibitem{roat02}
G. Roati, F. Riboli, G. Modungo, and M. Inguscio, Phys. Rev. Lett.
{\bf 89},  150403  (2002).

\bibitem{Loft01}
T. Loftus, C.~A. Regal, C. Ticknor, J.~L. Bohn, and D.~S. Jin,
Phys. Rev. Lett.
  {\bf 88},  173201  (2002).

\bibitem{Houb98}
M. Houbiers, H.~T.~C. Stoof, W.~I. McAlexander, and R.~G. Hulet,
Phys. Rev. A {\bf 57},  R1497  (1998).\\ See also ref. [23].

\bibitem{Hara02}
K.~M. O'Hara, S.~L. Hemmer, S.~R. Granade, M.~E. Gehm, J.~E.
Thomas, V.
  Venturi, E. Tiesinga, and C.~J. Williams, cond-mat/0207717  (2002).

\bibitem{Sten99}
J. Stenger, S. Inouye, M.~R. Andrews, H.-J. Miesner, D.~M.
Stamper-Kurn, and W.
  Ketterle, Phys. Rev. Lett. {\bf 82},  2422  (1999).

\bibitem{Robe00}
J.~L. Roberts, N.~R. Claussen, S.~L. Cornish, and C.~E. Wieman,
Phys. Rev.
  Lett. {\bf 85},  728  (2000).

\bibitem{Chin00}
C. Chin, V. Vuleti\'{c}, A.~J. Kerman, and S. Chu, Phys. Rev.
Lett. {\bf 85},
  2717  (2000).

\bibitem{Mewe96}
M.-O. Mewes, M.~R. Andrews, N.~J. van Druten, D.~M. Kurn, D.~S.
Durfee, and W.
  Ketterle, Phys. Rev. Lett. {\bf 77},  416  (1996).

\bibitem{Stoo88}
H.~T.~C. Stoof, J.~M.~V.~A. Koelman, and B.~J. Verhaar, Phys. Rev.
B {\bf 38},
  4688  (1988).

\bibitem{Kokk02}
Private communication with S. J. J. M. F. Kokkelmans and E.
Tiesinga.

\bibitem{Petr02}
Private communication with D. Petrov.

\bibitem{Houb99}
M. Houbiers and H.~T.~C. Stoof, Phys. Rev. A {\bf 59},  1556
(1999).

\bibitem{Timm01b}
E. Timmermans, Phys. Rev. Lett. {\bf 87},  240403  (2001).

\bibitem{Joch02}
M.~B. S.~Jochim, G. Hendl, J.~H. Denschlag, R. Grimm, A. Mosk, and
M.
  Weidem{\"u}ller, cond-mat/0207098  (2002).


\end{thebibliography}

\end{document}